# Cosmic rays studied with a hybrid high school detector array

A. Nigl[1], C. Timmermans[2], P. Schellart[1], J. Kuijpers[1], H. Falcke[1,3], A. Horneffer[1], C.M. de Vos[3], Y. Koopman[3], H.J. Pepping[3], and G. Schoonderbeek[3]

[1] Department of Astrophysics, IMAPP, Radboud University, Nijmegen, 6535 ED Nijmegen, The Netherlands
[2] NIKHEF and IMAPP, Radboud University, Nijmegen, 6535 ED Nijmegen, The Netherlands
[3] ASTRON, 7990 AA Dwingeloo, The Netherlands

**Summary**
The LORUN/NAHSA system is a pathfinder for hybrid cosmic ray research combined with education and outreach in the field of astro-particle physics. Particle detectors and radio antennae were mainly setup by students and placed on public buildings. After fully digital data acquisition, coincidence detections were selected. Three candidate events confirmed a working prototype, which can be multiplied to extend further particle detector arrays on high schools.

**Cosmic rays**
"Cosmic rays" is the generic name for different types of particles that hit the Earth's atmosphere with about the speed of light, varying in energy over a wide range. It was found that with increasing energy the number of cosmic rays arriving at the Earth drops sharply, see the flux-power spectrum in Fig. 1. The cosmic ray (CR) composition consists of protons (86%), $\alpha$-particles (11%), nuclei of heavier elements (1%), electrons (2%) and neutrinos (<1%) [1]. Most cosmic rays up to Gev energies are thought to be produced in our galaxy by stars and beyond these energies in supernova remnants (SNR); however the origin of the highest energetic particles is still unclear. An upper limit to the energy has been observed around 100 EeV (GZK cutoff [2], [3]). A uniform distribution on the sky strongly indicates that the highest energy CRs are produced in a variety of distant extra-galactic sources. Source candidates are collimated relativistic matter streams (jets) associated with galaxies emitting powerful radio emission (radio galaxies), as well as radio-loud young galaxies called quasi-stellar objects (quasars). Furthermore, bursts of high frequency electromagnetic emission called gamma ray bursts (GRBs) and magnetars, which are very compact stars with strong magnetic fields, are also candidates for the origin of the high energy CRs since they may be capable of the extreme particle acceleration required.

In the Earth's atmosphere these CRs initiate a cascade of particle collisions in which a large multiplicity of secondary particles of all kinds are produced. The creation and annihilation of particles in the cascade establishes a so-called air shower pancake traveling through the atmosphere. Additionally, the charged particles in the pancake - mainly the electrons and their counterparts the positrons - get deflected in the Earth's magnetic field and emit beamed coherent radiation called geosynchrotron emission [4]. The number of particles in the air shower and the peak voltage of the radio emission detectable on the ground depend approximately linearly on the energy of the primary particle. The radio emission increases with increasing angle between the cosmic ray trajectory and the Earth's magnetic field [5].

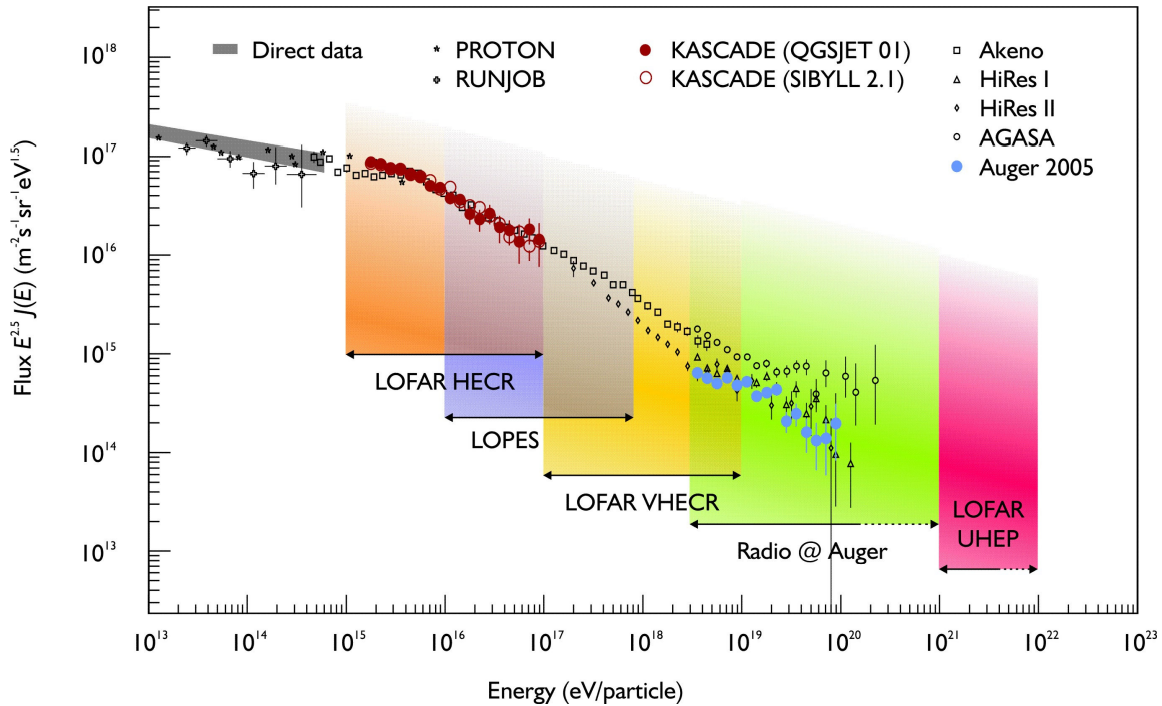

**Fig. 1:** Cosmic ray energy flux spectrum. The flux has been multiplied by a factor of $E^{2.5}$. Picturing the spectrum as an extended leg, the so-called knee and ankle are clearly visible at energies of about 3 PeV and 4 EeV, respectively. The very high energy cosmic rays (VHECR) are of interest for NAHSA and LORUN, which are sensitive to energies beyond 100 PeV (PeV = $10^{15}$ eV; EeV = $10^{18}$ eV).

**Why air showers?**
Every second, one low energy cosmic ray particle is hitting every square meter of the Earth's atmosphere. The resulting shower is attenuated in the atmosphere before reaching the ground and can, however, only be detected directly by a detector on a plane or balloon. The latter is the way cosmic rays have been discovered in 1912 by Victor F. Hess [6]. High energy cosmic rays hit an area of a square kilometer only once per year, thus requiring an expensive detector covering a large area for an effective measurement. On the other hand, from the air shower initiated by the primary particle, several hundred thousands of particles reach the ground within a radius of a few hundred meters, depending on the energy of the cosmic ray. Therefore, analyzing the shower to infer information about the cosmic ray particle is much more efficient. The largest cosmic ray detector, the Pierre Auger Observatory (www.auger.org) is currently under construction in Argentina, South America, and it is named after the discoverer of particle air showers [7]. This instrument detects the air showers mainly with particle detectors covering an area of 3000 km$^2$. The number of particles measured by the particle detectors allows one to estimate the energy of the primary particle, the direction of origin, and to find the lateral extension from a shower core, which depends on the species of the primary particle [8]. Next to particle detectors, four air fluorescence detectors measure the fluorescence light yield of about 10% of all detected showers.

In The Netherlands, ASTRON is building the largest low frequency radio telescope, which is called LOw Frequency ARray (LOFAR, www.lofar.org) and will be able to detect the radio emission from air showers with about 5000 small radio antennae grouped into about hundred stations of 200x200 m$^2$. Currently three test stations for LOFAR exist, two of which are optimized for cosmic ray air shower measurements. The first is the LOfar PrototypE Station (LOPES, www.lopes-project.org), which is made up of 30 simple dipole antennae to measure radio emission from all directions and became operational in March 2003. LOPES is triggered by particle detectors of the KArlsruhe Shower Core and Array Detector (KASKADE, www-

ik.fzk.de/KASCADE_home.html) situated on the Forschungszentrum Karlsruhe in Germany. The second prototype was built in December 2004, it is called LOfar at Radboud University Nijmegen (LORUN), and it is made up of four (crossed) dipole antennae on top of the building of the Radboud University Nijmegen (see Fig. 2). LORUN is triggered by two particle detectors, which are part of the High School Project on Astrophysics Research with Cosmics (HiSPARC). The pulse detected in the electric field of each antenna carries information about the longitudinal development of the shower such as the atmospheric depth at which the shower emission was maximal. This emission maximum provides a measure to distinguish between different species of primary particles. Furthermore, the radio antennae can be used as an interferometer to improve the accuracy in determining the direction from were the cosmic ray came [9].

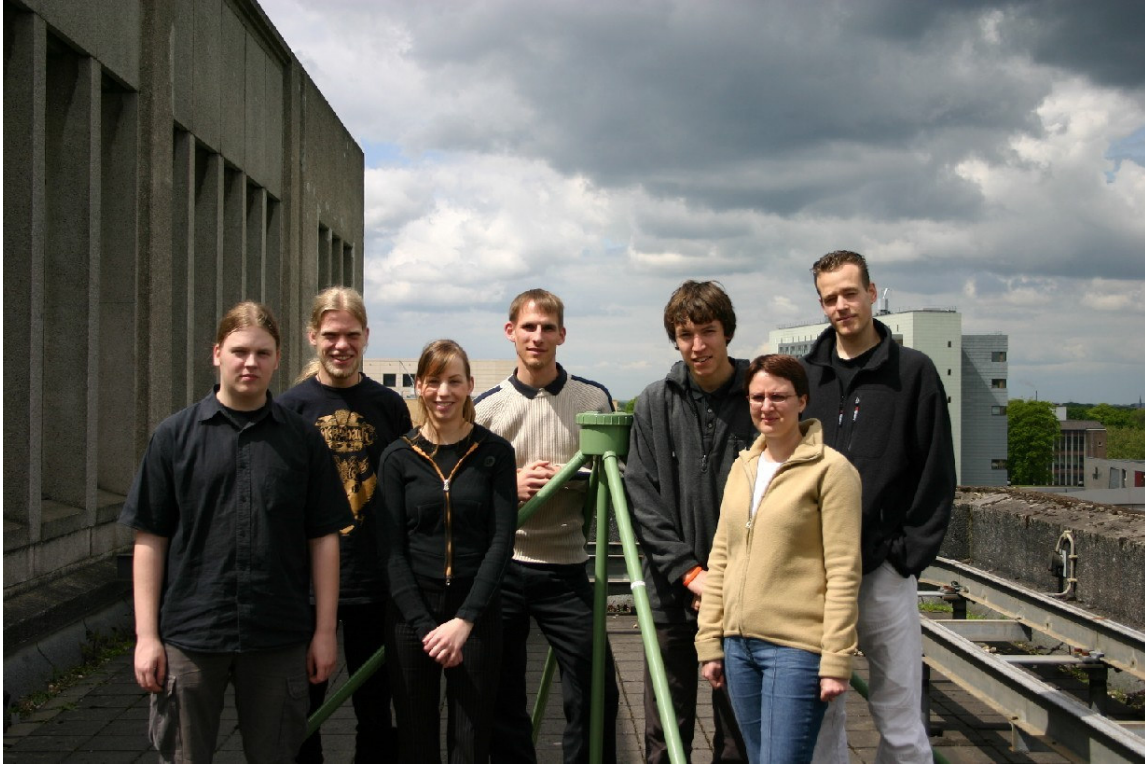

**Fig. 2:** LORUN antenna and the team of students that have been working on it on top of the building of the Radboud University Nijmegen (from left to right: Stefan Jansen, Karel Kok, Andra Verstee, Andreas Nigl, Jaap Kroes, Sandra Petrovic and Pim Schellart). The green LOFAR prototype antenna in the middle accommodates two (crossed) inverted-v-shaped dipoles in the legs of the PVC structure connected to two low-noise amplifiers (LNAs) located in the top box. The middle pole contains two cables for power and two for signal from the LNAs to the receivers.

**High school arrays**
Projects such as HiSPARC (www.hisparc.nl), the California HIgh school Cosmic ray ObServatory (CHICOS, www.chicos.caltech.edu) or the Stockholm Educational Air Shower Array (SEASA, neutrino.phys.washington.edu/~walta) place particle detectors on top of buildings such as, for example, schools and universities. Those particle detector arrays are scientifically interesting, since with the large area they cover, they contribute to the world record number of detected ultra high energy cosmic rays (UHECR). Furthermore, the high school arrays could be sensitive to the Zatsepin–Gerasimova effect [10, 11], in which highly energetic nuclei disintegrate in interactions with solar photons. As a result two simultaneous air showers could be observed on Earth. The confirmation of this effect would prove that a part of the high energetic radiation originates from interacting nuclei.

The HiSPARC project is made up of five clusters located in five Dutch cities. One cluster is the Nijmegen Area High School Array (NAHSA, hisparc.hef.kun.nl) in Nijmegen, which has been taking data since June 2002 and provides the triggers for the LORUN radio antennae. The other clusters are on buildings in Amsterdam, Leiden, Utrecht and Groningen. In the scope of HiSPARC, high school students receive hands-on training about how to build a particle detector and learn about its scientific applications. As a result, students extend the array and gain access to astro-particle physics and scientific instruments.

**HiSPARC/LORUN Setup**
Two particle detectors and four radio antennae have been placed on the roof of the building of the Faculty of Natural Sciences of the Radboud University, Nijmegen. The radio antennae and the data acquisition hardware for LORUN were provided by the Netherlands Foundation for Research in Astronomy (ASTRON, www.astron.nl). Each LORUN antenna accommodates two crossed dipoles, which are facing azimuth directions of 330° and 60°. Each dipole's signal is amplified and pre-filtered by a low-noise amplifier (LNA), further filtered by a band pass filter for the frequency band from 40 to 80 MHz and then digitized. The high performance Analog-Digital Converters (ADCs) combine a dynamic range of 12 bit with a rate of 80 million samples per second, which translates to a time resolution of 12.5 ns and a data rate of ~150 MB per second per channel (dipole). Because of the inverted-V-shape design of the antenna dipoles, the whole sky is instantly measured with a primary beam of 90° centered on the zenith.

The LORUN antennae are located near two scintillator-particle detectors of NAHSA accommodated in ski-boxes (see Fig. 3). NAHSA is made up of in total 8 twin detector stations of 0.5 $m^2$ each, located on high school buildings all over Nijmegen, with distances between 500 meters and 5 kilometers (see Fig. 4). More details on NAHSA can be found in [12].

The LORUN/HiSPARC hybrid system is triggered by a coincidence detection of two HiSPARC scintillators within a 1 µs time window for simultaneous data acquisition (DAQ) of both systems (see Fig. 5). The trigger rate counts up to more than 1000 events per day. HiSPARC measures traces of the current produced in the photo multiplier tube (PMT), which is generated by the amount of light produced in the detector by the shower particles. Each scope trace lasts 5 µs with a time resolution of 20 ns and is stored together with a Global Positioning System (GPS) timestamp. The LORUN system stores about half a millisecond of time series data before and after the trigger of each dipole with a time resolution of 12.5 ns. The read out and storage of the radio data takes about one second during which LORUN does not accept another trigger. The PC clock of LORUN is synchronized with the PC of the particle detector, using a network time protocol (NTP) to an accuracy of a few microseconds, so that the triggered events can be matched.

The triggered events of both experiments are matched off-line at the end of each day. Only events that were triggered by two NAHSA twin-detector stations are combined to exclude false-detections. About 10 events are obtained per day with energies beyond 10 PeV, corresponding to the sensitivity of this setup. The software scripts which perform the matching and storing of the recorded radio data are described in [12].

The radio data of the ten events per day are processed for coincidences of pulses detected by the eight radio dipoles around the trigger time. The shower emission wave-front crosses the detector from a certain direction and thus does not arrive at all antennae at the same time. Therefore the search for the pulses is performed in a time window determined by the light travel time across the maximum distance of all antennae pairs of 50 meters (170 ns). Unfortunately, there is environmental man-made radio frequency interference (RFI) produced by radio, TV and mobile phone broadcasting, which is narrow band, raises the noise level and makes the identification of pulses difficult. Therefore, the time-series from the dipoles are Fourier transformed and a sub-band from 45 MHz to 60 MHz with little RFI is chosen. The remaining RFI is down-weighted to reduce its contribution to the spectral power. After this digital filtering the spectrum is inverse Fourier transformed to obtain the air shower pulse. The off-line data analysis so described is performed with a graphical user interface of the LOPES-Tools, which provides FFT, filtering and beam-forming. For more details see the manual for the software package in [13].

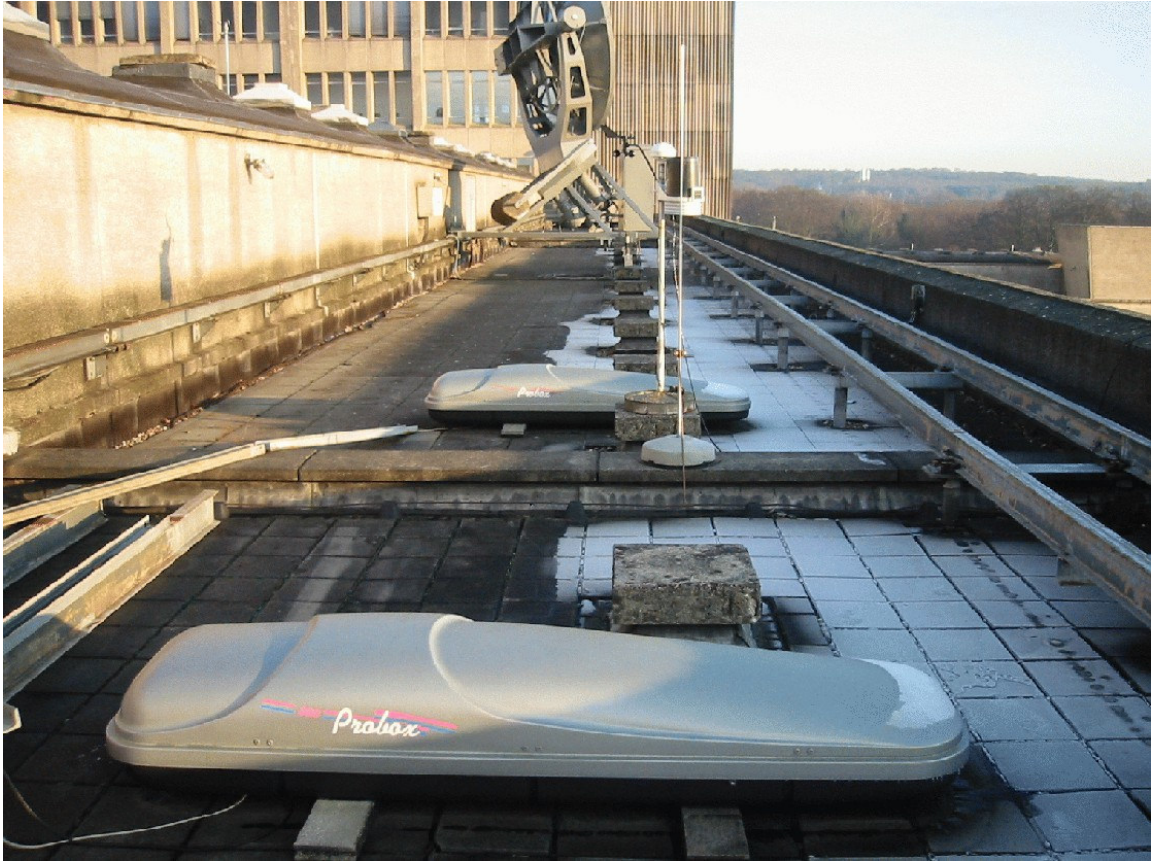

**Fig. 3:** Twin particle detector station on top of the building of the Radboud University Nijmegen. The ski boxes accommodate scintillator plates connected to a photo multiplier tube (PMT). The two poles between the ski boxes carry a weather station and the GPS for timing information. An analog signal cable runs from each box to the coincidence electronics.

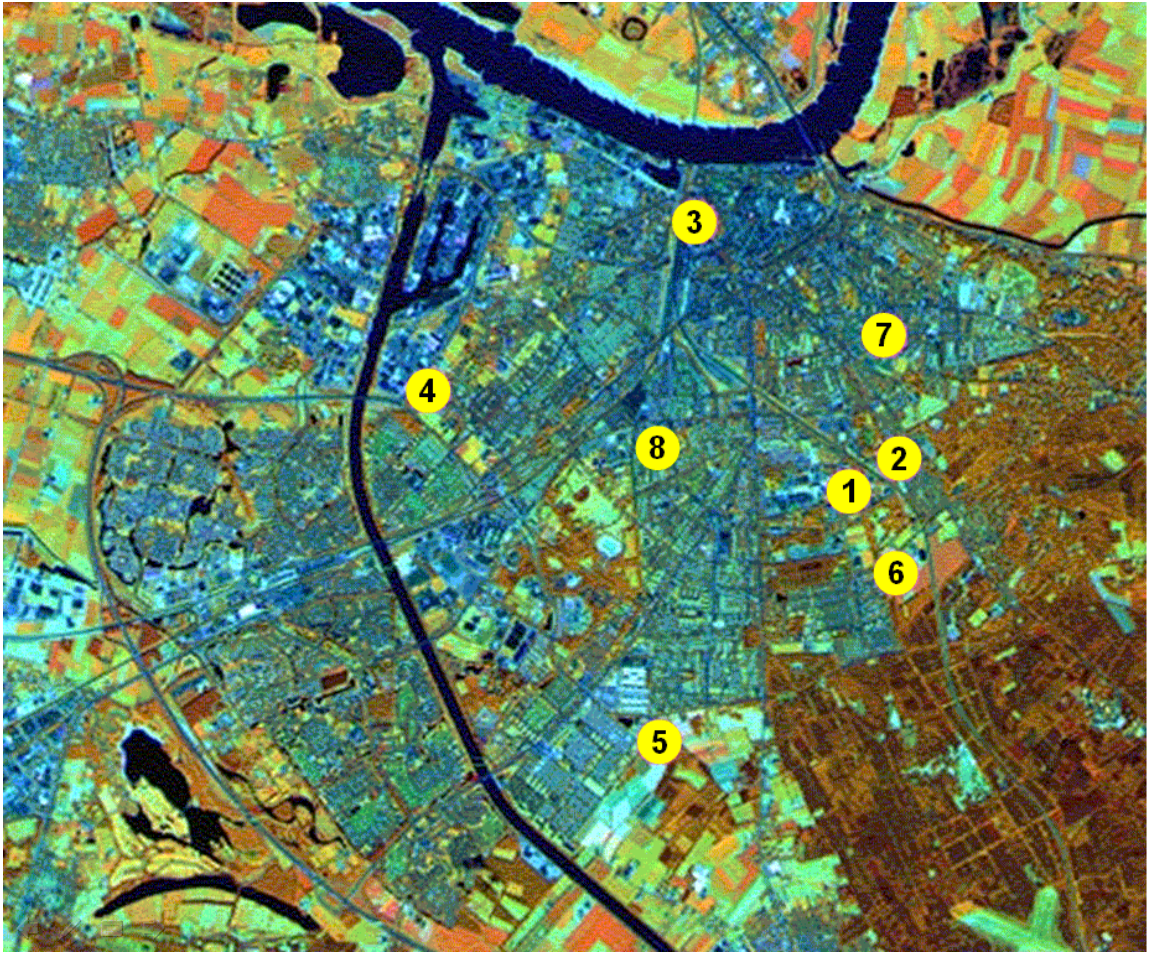

**Fig. 4:** Nijmegen with the locations of the NAHSA twin detectors and LORUN at position 1 on top of the building of the Faculty of Natural Sciences of the Radboud University.

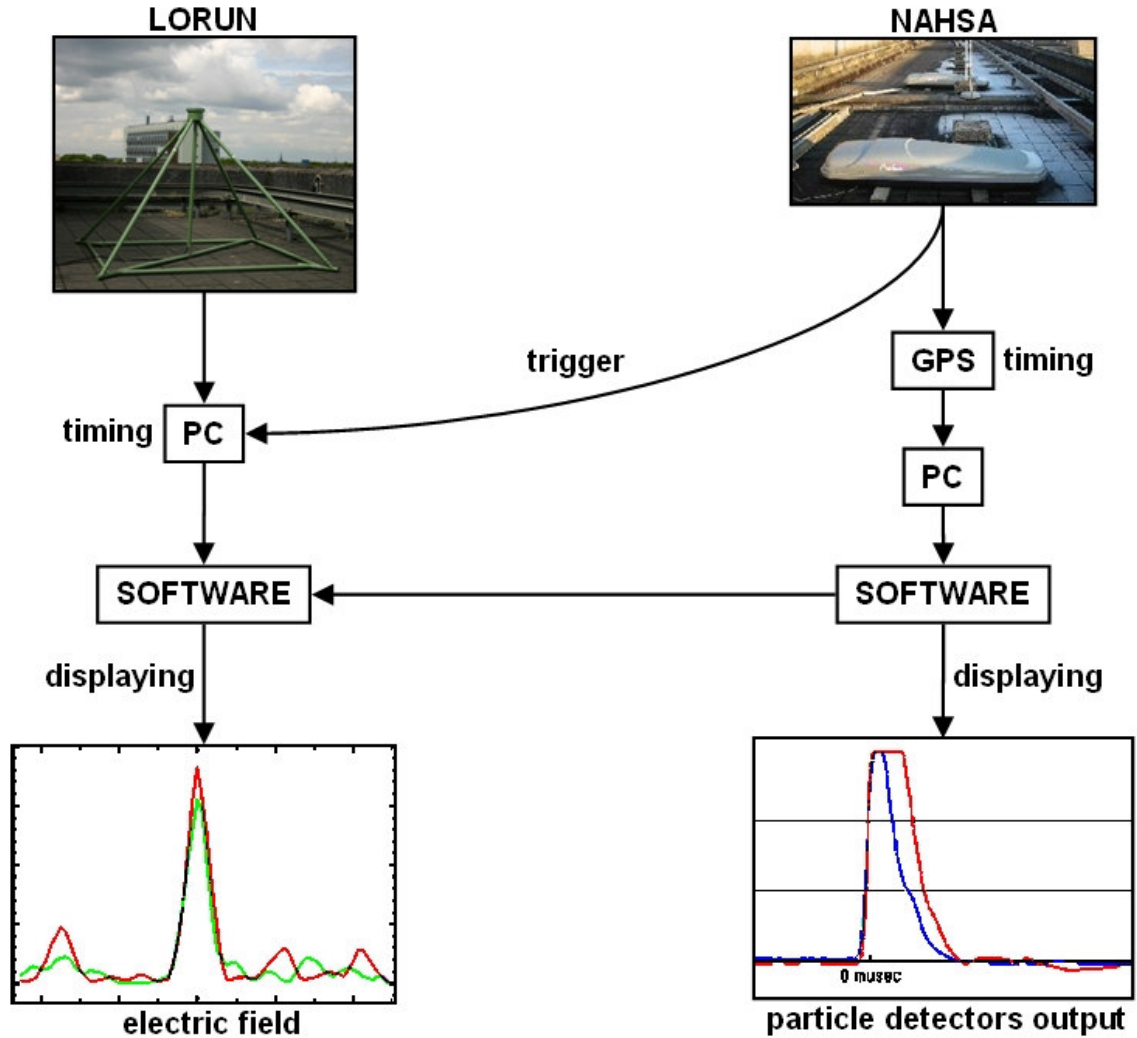

**Fig. 5:** LORUN/HiSPARC setup. LORUN is triggered by a coincidence detection of the two HiSPARC scintillators (top right photograph) within 1 ms. More than 1000 events are triggered per day. HiSPARC stores oscilloscope traces (bottom right plot) of both scintillator plates combined with a GPS timestamp. Each scope trace lasts 5 µs with a time resolution of 20 ns. The LORUN system stores about half a millisecond of time series data before and after the trigger of each dipole with a time resolution of 12.5 ns (bottom left plot). A LORUN antenna accommodating two dipoles is shown in the top left photograph.

**First Results**
The triggered and matched events were analyzed for coincident pulses in the measured electric field of the dipoles. Three candidates were found in the period from 2005-01-01 to 2005-08-01. HiSPARC detected all three events in two twin particle detectors, which are labeled with the station numbers 1 and 2 in the NAHSA layout in Fig. 4. Those two particle detector stations are spaced by 500 meters. LORUN measured these events with up to three operational dipoles. All three events were estimated to have a primary energy of larger than 0.2 EeV by the HiSPARC analysis software.
The first cosmic ray candidate was detected by two crossed dipoles, which are accommodated in the same antenna structure. Both antennae showed nice pulses of 50 ns length at the delay-corrected trigger time. The energy of the primary particle was estimated at more than 0.2 EeV.

The second cosmic ray candidate was detected in two radio antennae spaced by 30 meters. The relative peak height detected in the two parallel dipoles is proportional to the relative particle counts of HiSPARC (see Fig. 6). The difference in arrival time between the radio pulses is consistent with the time delay in arrival of detected particles at the two twin stations of NAHSA. The NAHSA detectors are spaced by 500 meters on a baseline nearly parallel to the LORUN antennae, which results in an estimated zenith angle of minimum 30°. The measured inclination suggests a strong primary particle, since the particle air shower has to reach the detector on a longer inclined trajectory through the Earth's atmosphere, without being attenuated. The energy was estimated at more than 0.3 EeV.

The third cosmic ray candidate was detected with three dipoles. The same two parallel dipoles of the previous event detected a much stronger radio pulse than a third perpendicular dipole, which is facing an azimuth direction of 330° and is part of the antenna structure of one of the parallel dipoles. Thus, the signal received in the North-South direction was much smaller than the signal received in the East-West direction. This result is in agreement with a geomagnetic emission mechanism, in which the Lorentz force deflects the charged air shower components in an East-West direction. The zenith angle was estimated at a minimum of 10° and the primary particle energy at more than 0.2 EeV.

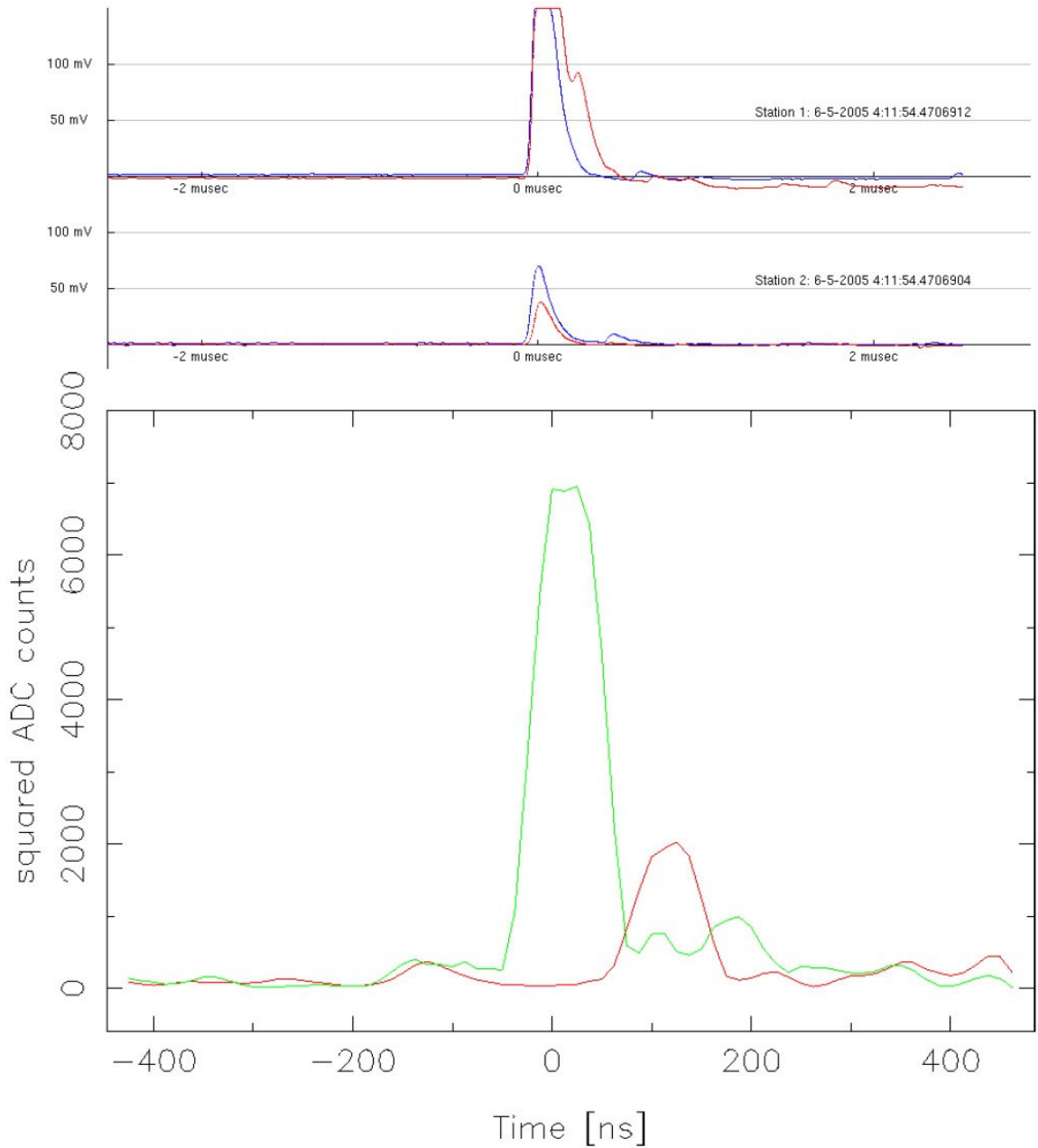

**Fig. 6:** Second cosmic ray candidate detected in two LORUN antennae spaced by 30 meters and two NAHSA twin detector stations spaced by 500 meters. BOTTOM: One dipole detected the right small peak (red) and a parallel one detected the left large peak (green). TOP: The relative peak height is proportional to the relative particle counts. The difference in arrival time at the radio dipoles is consistent with the time delay in arrival of particles at the NAHSA stations on a baseline parallel to the LORUN antennae, which results in an estimated zenith angle of 30˚. The measured inclination suggests a strong primary particle, since the particle air shower has to reach the detector on a longer inclined trajectory through the Earth's atmosphere, without being attenuated. The energy was estimated at more than 0.3 EeV.

**Outlook**
After setting up the hardware for the LOFAR prototype LORUN, it was straightforward to combine it with HiSPARC in a pathfinder experiment to detect radio emission and particles of cosmic ray air showers in coincidence. The NAHSA/LORUN experiment has been setup with the efforts of several undergraduate students. The HiSPARC experiment NAHSA in Nijmegen can benefit from the synergies of a hybrid detection of cosmic ray air showers in the radio regime combined with particle counts. The radio antennae have a high duty cycle and provide an independent way to determine the origin and primary energy of cosmic rays. The most powerful aspect of the radio detection is that the received signal integrates the whole longitudinal shower development, whereas particle detectors only receive those shower particles which arrive at ground level.

This pathfinder experiment has planted the seed to increase the outreach of cosmic ray science and the LOFAR project, with the opportunity for high school students to extend NAHSA and further HiSPARC stations in The Netherlands by adding radio antennae. However, additional development is needed in order to make the complete setup available for high schools. NIKHEF is developing a new readout for HiSPARC, as well as a readout scheme for radio detection of cosmic rays at Auger. Both of these developments will soon make a hybrid setup at high schools a reality.

**Biography**

Andreas Nigl is a mechanical engineer and he is doing his PhD research studies on cosmic rays at the Radboud University, Nijmegen in the scope of the LOFAR project. He is part of the LOPES group and collaborating with scientists of the Pierre Auger Observatory working in Nijmegen and at the Forschungszentrum Karlsruhe.

Charles Timmermans is a physicist working for NIKHEF at the Radboud University, Nijmegen. He is part of the Pierre Auger Collaboration, studying radio detection of cosmic rays at the highest energies, and he is a recipient of the 2007 EPS Outreach Prize.

Acknowledgments go to the team of students who helped to set up the experiment, to scientists and engineers from ASTRON, to Cees Brouwer and Peter Dolron of IMAPP and to the TechnoCenter of the Radboud University, Nijmegen.